\documentclass[a4paper, amsfonts, amssymb, amsmath, reprint, showkeys, nofootinbib, twoside]{revtex4-2}
\usepackage[english]{babel}

\usepackage[utf8]{inputenc}
\usepackage[colorinlistoftodos, color=green!40, prependcaption]{todonotes}
\usepackage{amsthm}
\usepackage{mathtools}
\usepackage{physics}
\usepackage{xcolor}
\usepackage{graphicx}
\usepackage[left=23mm,right=13mm,top=35mm,columnsep=15pt]{geometry} 
\usepackage{adjustbox}
\usepackage{placeins}
\usepackage[T1]{fontenc}
\usepackage{lipsum}
\usepackage{csquotes}

\usepackage{array}   
\newcolumntype{C}{>{$}c<{$}} 
\usepackage{subcaption}
\usepackage{comment}
\usepackage[pdftex, pdftitle={Article}, pdfauthor={Author}]{hyperref} 

\begin{document}

\title{Boosting Biomolecular Switch Efficiency With Quantum Coherence}

\author{Mattheus Burkhard}
       
    \affiliation{Clarendon Laboratory, University of Oxford, Parks Road, Oxford OX1 3PU, UK}
    \affiliation{Département de Physique, École Normale Supérieure Paris-Saclay, 4 Av. des Sciences, 91190 Gif-sur-Yvette, France}

\author{Onur Pusuluk}

    \affiliation{Faculty of Engineering and Natural Sciences, Kadir Has University, Fatih, Istanbul, Turkiye}

\author{Tristan Farrow}
       
    \email[Correspondence email address:] {tristan.farrow@physics.ox.ac.uk}
    \affiliation{Clarendon Laboratory, University of Oxford, Parks Road, Oxford OX1 3PU, UK}
    

\begin{abstract}

The resource theory of quantum thermodynamics has emerged as a powerful tool for exploring the out-of-equilibrium dynamics of microscopic and highly correlated systems. Recently, it has been employed in photoisomerization, a mechanism facilitating vision through the isomerism of the photo receptor protein rhodopsin, to elucidate the fundamental limits of efficiency inherent in this physical process. Limited attention has been given to the impact of energetic quantum coherences in this process, as these coherences do not influence the energy level populations within an individual molecule subjected to thermal operations. However, a specific type of energetic quantum coherences can impact the energy level populations in the scenario involving two or more molecules. In this study, we examine the case of two molecules undergoing photoisomerization to show that energetic quantum coherence can function as a resource that amplifies the efficiency of photoisomerization. These insights offer evidence for the role of energetic quantum coherence as a key resource in the realm of quantum thermodynamics at mesoscopic scales.

\end{abstract}


\maketitle

\section{Introduction}
\label{sec:intro}

Thermodynamics and quantum mechanics represent distinct disciplines that grapple with discrepancies between their fundamental principles. The emerging field of quantum thermodynamics~\cite{goold_role_2016, Vinjanampathy2016, QTD_2018, QTD_2019} bridges these two disciplines by proposing innovative strategies to reconcile them. One of the main discrepancies lies in the interpretation of energy within these two domains. The first law of thermodynamics traditionally characterizes an additive decomposition of energy into work and heat. In quantum mechanics, however, while energy is a measurable quantity, there are no directly observable counterparts for heat and work. By adopting a dynamical approach~\cite{Alicki2018} based on open quantum systems theory~\cite{BreuerAndPetruccione-2002}, quantum thermodynamics can precisely define and quantify heat and work within quantum systems.

Thermodynamic quantities such as free energy are well-defined only in equilibrium conditions, and in the limit of identical and independent distributions the conventional laws of thermodynamics hold true. Nonetheless, the intricate nature of quantum systems complicates matters. Unlike classical systems, we can only clone quantum systems with prior knowledge~\cite{Wootters1982}, and the presence of quantum coherence often pushes these systems far from equilibrium. Quantum correlations complicate the picture further by preventing quantum systems from becoming independent. To tackle these steep challenges, quantum thermodynamics employs information-theoretic approaches~\cite{2015_PhysReport_Nicole_Review, ng_resource_2018, lostaglio_introductory_2019, TORUN2023} falling under the framework known as quantum resource theories~\cite{coecke_mathematical_2016,chitambar_quantum_2019}. These approaches differentiate between states that are accessible by thermodynamical processes and those that are not.

The insight that heat and quantum coherence are convertible represents a significant advance in the realm of quantum thermodynamics~\cite{2015_NewJPhys_Brunner_T2Ent, 2017_PRA_Ozgur, 2018_PRE_RoleOfQCohInHT, 2019_PRE_T2QCoh, 2020_PRA_Brunner_AutonomousMultipartiteEntEngines, 2020_AsliReview}. Specifically, a form of energetic quantum coherence responsible for generating 
heat arises through the dynamical framework rooted in open quantum systems theory~\cite{2016_Entropy_Ozgur, 2019_PRE_Ozgur_Multiatom, AHF_2019_PhysRevResearch_Petruccione, AutoThermalMach_2019, 2020_PRA_HorizantolCohAndPops}. This specific manifestation of coherence is characterized by superpositions within degenerate energy states and has led to various nomenclatures such as ``heat-exchange coherence,''~\cite{2016_Entropy_Ozgur, 2019_PRE_Ozgur_Multiatom} ``internal coherence,''~\cite{AHF_2019_PhysRevResearch_Petruccione} or ``horizontal coherence''~\cite{AutoThermalMach_2019, 2020_PRA_HorizantolCohAndPops}. This kind of coherence has enabled the establishment of a quantum Onsager relation that links coherence flow and heat flow~\cite{pusuluk_quantum_2021}.

The information-theoretical approach refers to the same type of coherence as ``zero-mode coherence.'' In this framework~\cite{lostaglio_quantum_2015,biswas_fluctuation-dissipation_2022}, one considers zero-mode coherences in conjunction with the energy level populations within the thermomajorization criterion (an extension of the second law of thermodynamics~\cite{lostaglio_introductory_2019}). 
While the study of coherence in thermal processes within the framework of open quantum systems has garnered significant attention, by comparison, the study of their effects within resource theories remains relatively uncharted territory. Our work aims to fill this gap.

We consider the process of photoisomerization to illustrate how these principles can usefully be applied. Photoisomerization - or photoswitching - stands as an example of the explanatory capacity of the thermomajorization criterion in elucidating the behaviour of systems far from equilibrium such as biological molecules~\cite{yunger_halpern_fundamental_2020}. The same model system has also been used to identify contributions from non-Markovianity in thermomajorization~\cite{spaventa_capacity_2022}. These prior investigations focus on a single photoswitching molecule where the contributions of coherence to the efficiency of photoisomerization cannot be examined since in a single-molecule system the zero-mode quantum coherence is absent. Here, we explore a scenario where two identical rhodopsin molecules are simultaneously stimulated by a single photon, leading to the sharing of zero-mode coherence between them and giving rise to nontrivial contributions of coherence to the final state of the reaction.


\section{Resource theory}
\label{sec:resource theory}
\subsubsection{Defining possible operations}

In quantum thermodynamics, free operations at inverse temperature $\beta=1/k_BT$ are called \textit{thermal operations}. These operations do not require any additional resources (no external `battery') and can be the following~\cite{lostaglio_introductory_2019, chitambar_quantum_2019}:
\begin{itemize}
    \item[(i)] Contact with a thermal bath B (free use of states with a density matrix $\rho_B = \frac{e^{-\beta H_B}}{Z}$).
    \item[(ii)] Any energy-preserving global unitary transformation $U$ on the whole system.
    \item[(iii)] Tracing out any subsystem, and in particular the bath B.
\end{itemize}

This means that the thermal operation from a certain initial state $\rho$ to a final state $\sigma$ can be described by the functional~$\mathcal{T}$:
\begin{equation}
    \mathcal{T}[\rho]=\mathrm{Tr}_B\left[U(\rho\otimes\rho_B)U^\dagger\right]=\sigma
\end{equation}
In other words a thermal operation on a system $\rho$ is any process that only takes thermal energy from the bath and preserves the total energy.

\subsubsection{Constructing a Lorenz curve}
Now, let us define thermomajorization. It is a mathematical tool that allows us to determine the relative ordering of states based on their energy distributions. When one thermal state thermomajorizes another, it means that the former has a more organized and concentrated energy distribution compared to the latter. It resembles the second law of thermodynamics where organized systems have lower entropy than those where the energy is distributed over all degrees of freedom. The thermomajorization relation is important in resource theory as it helps quantify the usefulness or value of states for performing certain thermodynamic tasks.

$H=\sum_{j} E_j \ketbra{j}{j}$ denotes the Hamiltonian that governs the mechanics of our system. Its state is represented by a density matrix $\rho$ whose diagonal elements are $\rho_{jj}\ketbra{j}{j}$. The coefficients can be regrouped in what is called a population vector:
\begin{equation}
\begin{split}
    \vec{p}&=(\rho_{11},\rho_{22},...,\rho_{dd})\\
    &\equiv(p_1,p_2,...,p_d)
\end{split}
\end{equation}
It contains the probabilities to be in one of the energy eigenstates, with $\sum_j^dp_j=\Tr(\rho)=1$.
To explain thermomajorization we need to define a curve for this state called the Lorenz curve and denoted by $L(\Vec{p})$. We can construct this curve with a procedure that includes two main steps. First, one has to calculate the rescaled coefficients $p_je^{\beta E_j}$, and order them from greatest to least:
\begin{equation}
    p_{j'} e^{\beta E_{j'}} \leq p_{k'} e^{\beta E_{k'}} \text{, for all } j'  > k'
    \label{eq:reordering}
\end{equation}
Secondly, one has to consider the points:
\begin{equation}
    \left(\sum^\alpha_{j=1} e^{-\beta E_j}, \sum^\alpha_{j=1}p_j\right)\text{, with }\alpha = 1, 2,..., d
\end{equation}
Connecting them with straight lines beginning at the origin defines a piecewise-linear curve, the Lorenz curve. The x-coordinates run from 0 to Z, the partition function $Z=\sum_i^de^{-\beta E_j}$. The y-coordinates go from 0 to 1, the sum of all probabilities. 

Now let us say ($\sigma$, $H'$) defines another state with a population vector $\Vec{q}=(q_1,...,q_{n'})$. If the ($\rho$, $H$) curve lies above or on the ($\sigma$, $H'$) curve, then ($\rho$, $H$) is said to thermomajorize ($\sigma$, $H'$). This can be denoted by 
\begin{equation}
\begin{split}
    [L(\Vec{p})](x)&\geq [L(\Vec{q})](x), ~\forall x\in[0,Z]\\
    \text{Shorthand notation:~~}L(\Vec{p})&\geq L(\vec{q})
\end{split}
\end{equation}

A mathematical theorem \cite{lostaglio_introductory_2019,horodecki_fundamental_2013} links thermal operations to thermomajorization. If and only if ($\rho$, $H$) thermomajorizes ($\sigma$, $H'$) there exists some thermal operation $\mathcal{T}$ that maps $\rho$~to~$\sigma$. In other terms, if $\rho$ (resp. $\sigma$) has a population vector $\Vec{p}$ (resp. $\Vec{q}$), then:
\begin{equation}
    \Big(L(\Vec{p})\geq L(\vec{q})\Big)\Longleftrightarrow\Big(\exists \mathcal{T}, \mathcal{T}[\rho]=\sigma\Big)
\end{equation}

\subsubsection{Importance of off-diagonal elements (coherences)}
\label{sec:off-diagonal elmts}

The density matrix $\rho$ of a system can be written as the sum of its elements in the energy eigenbasis.
\begin{equation}
    \rho=\sum_{n,m}\rho_{nm}\ketbra{n}{m}
\end{equation}
where $n$ designates the state of energy $E_n=\hbar\omega_n$. The transition energy between two states $\omega=\omega_n-\omega_m$ defines a mode. All elements $\rho_{nm}$ with the same transition energy belong to the same \textit{$\omega$-mode}. 

\begin{equation}
    \rho^{(\omega)}\equiv\sum_{n,m|~\omega_n-\omega_m=\omega}\rho_{nm}\ketbra{n}{m}
\end{equation}
This way the density matrix can be decomposed in its $\omega$-modes.
\begin{equation}
    \rho=\sum_\omega\rho^{(\omega)}
\end{equation}
One can prove that these $\omega$-modes evolve independently using the time-translation symmetry of thermal operations. This symmetry is defined by (more on this in~\cite{chitambar_quantum_2019,lostaglio_introductory_2019}):
\begin{equation}
    \mathcal{T}[e^{-iHt/\hbar}\rho e^{iHt/\hbar}]=e^{-iHt/\hbar}\mathcal{T}[\rho]e^{iHt/\hbar}
    \label{eq:time-translation symmetry}
\end{equation}
The claim is that $\mathcal{T}[\rho^{(\omega)}]$ is also a $\omega$-mode.
\begin{equation}
    \begin{split}
        e^{-iHt/\hbar}\mathcal{T}[\rho^{(\omega)}]e^{iHt/\hbar}=&\mathcal{T}[e^{-iHt/\hbar}\rho^{(\omega)}e^{iHt/\hbar}]\\
        =&\mathcal{T}[e^{-i\omega t}\rho^{(\omega)}]\\
        =&e^{-i\omega t}\mathcal{T}[\rho^{(\omega)}]
    \end{split}
\end{equation}
This claim holds, which shows that all $\omega$-modes of a density matrix evolve independently during a thermal operation. For a non-degenerate system the only elements $\rho_{nm}$ where $\omega_n-\omega_m=0$ are the diagonal elements. This is why in certain cases we can neglect the impact of the off-diagonal elements in thermomajorization. 

In previous studies of photoisomerization using thermomajorization~\cite{yunger_halpern_fundamental_2020,spaventa_capacity_2022}, the model only involved a non-degenerate energy eigenbasis. This explains why it was not previously necessary to consider coherences in the calculations.

For a system with degenerate energies one has to include the off-diagonal elements in the thermomajorization procedure. The method described below was presented in \cite{biswas_fluctuation-dissipation_2022}.
$$\rho\xrightarrow{~~~\mathcal{U}~~~}\rho^*\xrightarrow{~~~\mathcal{T}~~~}\sigma^*\xrightarrow{~~~\mathcal{U^\dagger}~~~}\sigma$$
First one diagonalizes the initial density matrix $\rho$ with a unitary transformation $\mathcal{U}$. In fact $\rho$ restricted to its 0-modes is already block-diagonal in the energy-eigenbasis, so $\mathcal{U}$ will act on the blocks of degenerate energy subspaces. This transformation is allowed because energy-preserving unitaries are free and acting on subspaces of constant energy preserves the energy. We obtain $\rho^*$ where all the 0-modes are on the diagonal. So we have a new initial population vector $\Vec{p}_{i}^{~*}$ on which to apply the thermomajorization procedure $\mathcal{T}$. This gives us a final state $\sigma^*$ and we can apply the inverse transformation $\mathcal{U}^\dagger$ to switch back to the original basis and get our final result $\sigma$.
This is an innovative method for studying coherences and quantifying their effects on photoisomerization.

\section{Model of photoisomerization}
\label{sec:model}
\begin{figure}[h]
    \centering
    \includegraphics[height=4.5cm]{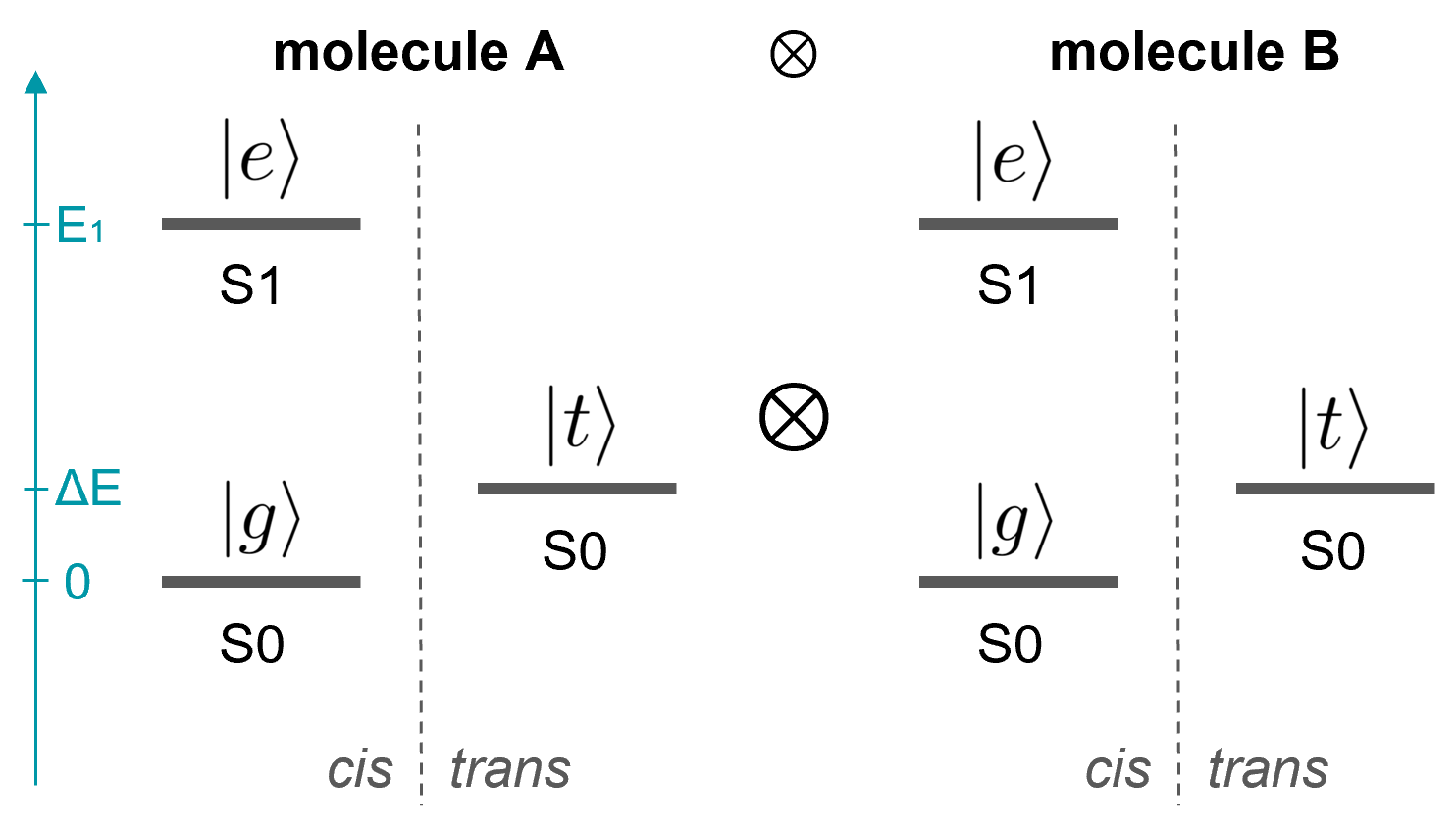}
    \caption{\textbf{A simplified picture of the energy levels of two molecules.} The full energy eigenbasis consists of ${\{\ket{i}\otimes\ket{j}\}}$ with $i,j\in\{e,g,t\}$.}
    \label{fig:cis-trans_2mol}
\end{figure}

Thermomajorization can give us the optimal quantum thermodynamical
yield of a given process.
One goal of this work is to complete our understanding of the quantum yield of a molecule undergoing photoisomerization. Also called photoswitching, it is triggered in certain molecules when they absorb a photon, the acquired energy induces a rotation around one of the molecule's carbon double bonds. As for example with the protein rhodopsin, the molecule switches from a \textit{cis} to \textit{trans} configuration when it absorbs a photon~\cite{ernst_microbial_2014,hahn_quantum-mechanical_2000,wang_vibrationally_1994}. Rhodopsin is a protein responsible for vision in human and animal retinas, by photoisomerizing it transforms the optical signal into a chemical chain reaction which transmits a signal to the brain. The resource-theoretical method for the case of a single molecule undergoing photoisomerization was developed in~\cite{yunger_halpern_fundamental_2020,spaventa_capacity_2022}. The total Hamiltonian presented in equation~(\ref{eq:total Hamiltonian}) is restricted to the evolution of the molecule along the reaction coordinate of the chemical transformation which is the angle of rotation $\phi$ around the double bond. Similar models have been developed for different biosystems~\cite{pusuluk_proton_2018}.
\begin{equation}
    H=\int_{\phi=0}^\pi H_{mol}(\phi)
    \label{eq:total Hamiltonian}
\end{equation}
In fact, by focusing on the initial and final state, $\phi=0$ or $\pi$, we can ignore the intermediate states and energy barriers when we want to obtain the optimal possible yield allowed by thermal operations. This is justified in the context of resource theory that answers the question of whether a certain transformation is possible or not. The theory makes general statements about complex dynamics and this applies to studying the efficiency of photoisomerization. The process can be seen as the evolution of an effective 3-level system. As shown in figure~\ref{fig:cis-trans_2mol} for one molecule (A or B) these states are called: $\left\{\ket{g};~\ket{e};~\ket{t}\right\}$. They correspond to the electronic \textit{ground} and \textit{excited} states of the molecular \textit{cis} configuration, and the \textit{trans} ground state. The omission of the \textit{trans} excited state can be justified by saying that it does not impact the final population in $\ket{t}$. If a molecule reaches the \textit{trans} excited state, it can freely relax into the corresponding ground state and emit its surplus of energy into the bath $B$.
The total molecular state is encoded by a density matrix whose diagonal elements form a population vector $\vec{p}=(p_g,p_e,p_t)$. It corresponds to the energy eigenbasis restricted to $\phi=0$ and $\pi$. The non-diagonal elements of the density matrix cannot influence the quantum yield in such a system, as explained in \cite{lostaglio_introductory_2019,chitambar_quantum_2019}. Indeed the energy eigenbasis is non-degenerate, with $\Delta E$ describing the energy gap from \textit{cis} to \textit{trans} and $E_1$ the energy of the photoexcitation (see fig.~\ref{fig:cis-trans_2mol}). Typical values in the case of photoisomerization of rhodopsin are $E_1=2.48$eV and $\Delta E=1.39$eV~\cite{wang_vibrationally_1994,hahn_quantum-mechanical_2000,johnson_primary_2017,pedraza-gonzalez_a-arm_2019}.
In this case the diagonal elements are the only 0-modes, so coherences do not affect their evolution.

When describing photoisomerization, the initial conditions correspond to the photoabsorption event. The initial population vector is a mixture of the \textit{cis} ground state $S_0$ and its first excited state $S_1$ with $p$ being the probability of photoexcitation.
\begin{equation}
    \vec{p_{i}}=\left(1-p,p,0\right)
    \label{eq:1mol init pop vec}
\end{equation}
The final state can be described by any combination $\vec{q}=(q_g,q_e,q_t)$ as long as there is a thermal operation that maps it to the initial one. That means that $\Vec{q}$ is thermomajorized by $\vec{p_{i}}$, or in mathematical terms $L(\Vec{p}_i)\geq L(\Vec{q})$.

Let us expand the model to N=2 molecules, because coherences will start to play a role due to degeneracy. 

The 3-level approximation leads to an overall dimension of $3^N=3^2=9$, which complicates calculations and increases computational cost.  

To define thermodynamic quantities like energy, heat, work, temperature, and so on locally, one should consider the eigenbasis of the sum of local Hamiltonians as the energy basis.
\begin{equation}
    H=H_A+H_B
\end{equation}

The pure state of the two molecules can be described as a sum of combinations of basis states ($g,~e~\text{and}~t$). 
\begin{align}
    &\ket{\psi}=\sum_{i,j\in\{g,e,t\}}\alpha_{ij}\ket{i}\otimes\ket{j}\\
    \label{eq:state describtion}
    \text{Or: }~
    &\ket{\psi}=\sum_{i,j\in\{g,e,t\}}\alpha_{ij}\ket{ij}
\end{align}
For thermomajorization, the initial state in the photoisomerization process is given by the population vector that describes the occupation probabilities in the \textit{cis} subspace $\{{gg},{ge},{eg},{ee}\}$. 
\begin{equation}
    \vec{p_i}=\left(p_{gg},p_{ge},p_{eg},p_{ee},0,0,0,0,0\right)
    \label{eq:initial p}
\end{equation}
where $\sum_{i,j\in\{g,e\}}p_{ij}=1$. As expected, the probabilities of having some molecule in $t$ are equal to $0$ before the photoisomerization. Afterwards the final state can be any $\vec{q}=\left(q_{gg},q_{ge},q_{eg},q_{ee},q_{gt},q_{tg},q_{et},q_{te},q_{tt}\right)$ as long as $L(\vec{p}_i)\geq L(\vec{q})$.
The optimal quantum yield can be defined as the sum of probabilities where any of the two molecules has reached the \textit{trans} state, which means that at least one of them is in $t$:
\begin{equation}
    QY_{\text{any}}\equiv\max_{L(\vec{p}_i)\geq L(\vec{q})}\sum_{i\in\{g,e\}}\left(q_{it}+q_{ti}\right)+q_{tt}
    \label{eq:QY}
\end{equation}
Another possible definition of the quantum yield is:
\begin{equation}
    QY_{\text{both}}\equiv \max_{L(\vec{p}_i)\geq L(\vec{q})}q_{tt}
\end{equation} when both have reached the \textit{trans} state.

For two molecules there is a degeneracy in the \textit{cis} subspace when one molecule is excited, $E_{ge}=E_{ge}=E_1$, two different states $\ket{ge}$ and $\ket{eg}$ correspond to this energy. 
The initial state does not have to be a pure state, so we can consider a mixed state, where the off-diagonal coefficients in the restricted matrix $\rho\vert_{E_1}$ are set by a parameter $\lambda$. 

\begin{equation}
    \rho\vert_{E_1}=
    \begin{pmatrix}
        |\alpha_{ge}|^2 & \lambda\\
        \lambda^* & |\alpha_{eg}|^2
    \end{pmatrix}\equiv
    \begin{pmatrix}
        p_{ge} & \lambda\\
        \lambda^* & p_{eg}
    \end{pmatrix}
\end{equation}

with $0\leq|\lambda|\leq \sqrt{p_{eg}p_{ge}}$. This parameter encodes the amount of decoherence the state has experienced. This is reflected in a loss of purity:
\begin{equation}
\Tr(\rho\vert_{E_1}^{~~2})=p_{ge}^2+2|\lambda|^2+p_{eg}^2\leq(p_{ge}+p_{eg})^2
\end{equation}
After diagonalizing this mixed state one obtains:
\begin{align}
    \rho^*\vert_{E_1}&=
    \begin{pmatrix}
        p_+ & 0\\
        0 & p_-
    \end{pmatrix}\\
    \text{with~~~~}
    p_\pm=\frac{|\alpha_{ge}|^2+|\alpha_{eg}|^2}{2}
    &\pm\sqrt{\left(\frac{|\alpha_{ge}|^2-|\alpha_{eg}|^2}{2}\right)^2+|\lambda|^2}
    \label{eq: p+ and p-}
\end{align}

This gives us an updated version of the probability vector introduced in equation~(\ref{eq:initial p}) in the presence of decoherence.
\begin{equation}
    \vec{p}_{i}^{~*}=\left(p_{gg},p_+,p_-,p_{ee},0,0,0,0,0\right)
    \label{eq:initial diagonalized p}
\end{equation}
The unitary conserves the global probability to be in the $E_1$ subspace $p_++p_-=p_{ge}+p_{eg}$.

\section{Results and discussion}
\label{sec:results}

One can imagine the following thought-experiment to study coherence effects. An incoming photon goes through a first beamsplitter that transmits only a portion $p$ of its wavefunction. That portion is then guided towards a perfect 50/50 beamsplitter and lands in a state of superposition across modes A and B. So there is a probability $1-p$ that the photon will be in neither A nor B, and a probability $p/2$ to be respectively in A or B exclusively. After the beamsplitter the photonic state is:
\begin{equation}
    \begin{split}
        \ket{\gamma}=~&\sqrt{1-p}\ket{A:0,B:0}\\
        +~& \sqrt{\frac{p}{2}}\Big(\ket{A:1,B:0}+\ket{A:0,B:1}\Big)
    \end{split}
\end{equation}
where $1$ and $0$ encode the presence and absence of the photon in the channel A or B. In each channel there is a photoisomerizable molecule, also respectively called A and B. Now, $p$ is - like in equation~(\ref{eq:1mol init pop vec}) - analogous to the probability that the molecule absorbs the photon, one obtains the following two-molecular state:
\begin{equation}
    \begin{split}
        \ket{\psi} = ~&\sqrt{1-p}\ket{A:g,B:g}\\
         +~ &\sqrt{\frac{p}{2}}\Big(\ket{A:e,B:g}+\ket{A:g,B:e}\Big)
    \end{split}
\end{equation}
The density matrix, restricted to its 0-modes, can be written in a simpler form as:
\begin{equation}
    \begin{split}
        \rho = & ~(1-p)\ketbra{gg}{gg}\\ + & ~\frac{p}{2}\left(\ket{ge}+\ket{eg}\right)\left(\bra{ge}+\bra{eg}\right)
    \end{split}
\end{equation}
Restricted to the initial subspace of excited and ground state of the \textit{cis} configuration, one obtains the following matrix representation:  
\begin{equation}
    \rho\vert_{e,g}=\begin{pmatrix}
1-p & 0 & 0 & 0\\
0 & p/2 & p/2  & 0\\
0 & p/2  & p/2  & 0\\
0 & 0 & 0 & 0
\end{pmatrix}
\end{equation}
There are now two off-diagonal which are also 0-modes appearing in the initial configuration. Their presence will impact the final quantum yield.

Now, if we assume that the process described so far is not taking place in a closed system, then some decoherence can take place. The off-diagonal elements will be subjected to it and their amplitude may vary. In particular we can replace the coherence with a parameter $\lambda$, such that:
\begin{equation}
\Tilde{\rho}\vert_{e,g}=\begin{pmatrix}
1-p & 0 & 0 & 0\\
0 & p/2  & \lambda & 0\\
0 & \lambda^* & p/2  & 0\\
0 & 0 & 0 & 0
\end{pmatrix}
\end{equation}
with $0\leq|\lambda|\leq p/2 $. When diagonalized:
\begin{equation}
\Tilde{\rho}^*\vert_{e,g}=\begin{pmatrix}
1-p & 0 & 0 & 0\\
0 & p/2 + |\lambda|  & 0 & 0\\
0 & 0 & p/2 - |\lambda| & 0\\
0 & 0 & 0 & 0
\end{pmatrix}
\end{equation}

This gives us the following initial population vector:
\begin{equation}
    \Vec{p}_{i}=\left\{1-p,p/2 + |\lambda|,p/2 - |\lambda|,0,0,0,0,0,0\right\}\equiv \Vec{p}_{sup}(p,\lambda)
    \label{eq:pop_vec-superposition}
\end{equation}
called $\vec{p}_{sup}$ for \textit{sup}erposition of molecular excitations.

Now we can study the influence of the coherence parameter $\lambda$ on the photoisomerization efficiency using thermomajorization.

For the following figures, the typical parameters for rhodopsin ($E_1=2.48$eV and $\Delta E=1.39$eV) were used. We also rescaled the inverse temperature to be ${\beta=1\text{eV}^{-1}}$. For the initial $\Vec{p}_{sup}$,  we used $p=0.7$ and compared low coherence ($\lambda=0.02$) and high coherence ($\lambda=0.2$) regimes.

\begin{figure}[h]
    \centering
    \includegraphics[width=\linewidth]{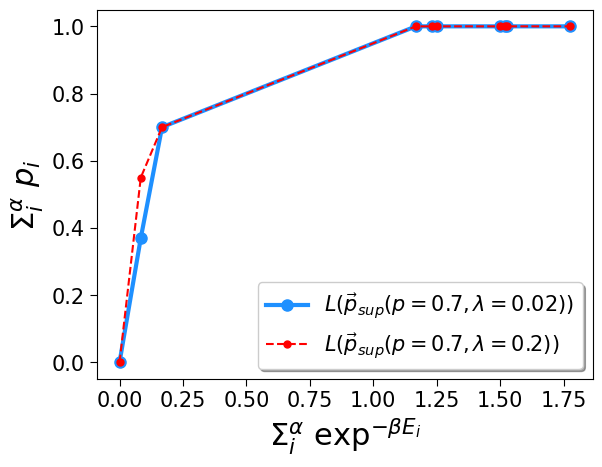}
    \caption{\textbf{Comparison of Lorenz curves of initial states with differing coherence}, but same $p=0.7$. High-coherence state $\Vec{p}_{sup}(\lambda=0.2)$ has a higher curve than low coherence $\Vec{p}_{sup}(\lambda=0.02)$. Coherence leads to higher thermomajorization potentiality.}
    \label{fig:lorenz_hl_vs_ll}
\end{figure}

As shown in figure~\ref{fig:lorenz_hl_vs_ll} the latter case has a slightly increased Lorenz curve in the range between 0 and 0.25 on the horizontal axis. This feature can be understood with the following argument. For the initial population vector $\Vec{p}_{sup}$, we have $p_\pm=p/2 \pm |\lambda|$. When constructing the Lorenz curve $L(\vec{p}_i)$, the population vector has to be reordered, and the resulting vector can be called $\vec{p_i}'$. For all values of $\lambda$, $p_+$ will be ordered before $p_-$, because $p_+e^{\beta E_1}>p_-e^{\beta E_1}$, and let us say for simplicity that they end up being adjacent at positions $k+1$ and $k+2$. This means that there are $k$ other terms in $\vec{p_i}'$ before $p_+$ and $p_-$. This is in fact the case in fig.~\ref{fig:lorenz_hl_vs_ll} and here $k=1$ (there is only one data point preceding $p_+$ and $p_-$, the origin (0,0)). Now, the point $k+1$ of the Lorenz curve will take the value given below on its y-axis.
\begin{equation}
    \sum_{j=1}^k(\vec{p_i}')_j+p_+=\sum_{j=1}^k(\vec{p_i}')_j+p/2 + |\lambda|
\end{equation}
The higher the coherence parameter $\lambda$, the higher the Lorenz curve $L(\vec{p}_{sup}(p,\lambda))$ at this point. The next data point, numbered $k+2$, has the y-coordinate:
\begin{equation}
    \sum_{j=1}^k(\vec{p_i}')_j+p_++p_-=\sum_{j=1}^k(\vec{p_i}')_j+p
\end{equation}
This result does not depend on $\lambda$, because we always have $p_++p_-=p_{ge}+p_{eg}$. So, whatever the value of $\lambda$, the Lorenz curve reaches the same height after $k+2$ points. Hence, by increasing $\lambda$ we obtain higher Lorenz curves with increasing convexity. This is shown by the red dotted line above the blue one in figure~\ref{fig:lorenz_hl_vs_ll}, ${L(\vec{p}_{sup}(p=0.7,\lambda=0.2))\geq L(\vec{p}_{sup}(p=0.7,\lambda=0.02))}$.
\begin{figure}[h]
    \begin{subfigure}{\linewidth}
    \includegraphics[width=\textwidth]{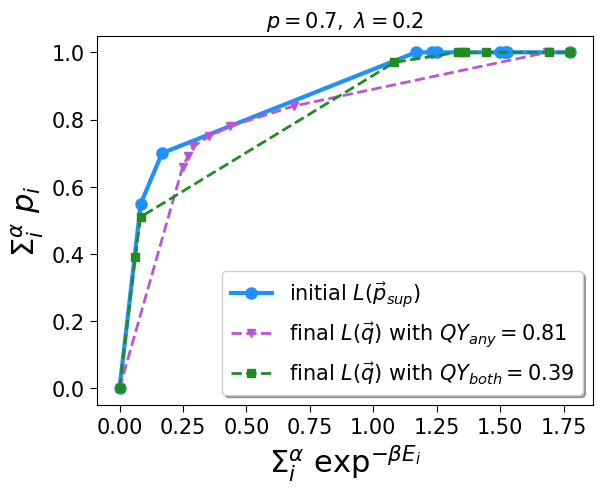}
    \caption{\textbf{High coherence}, with initial state: $\vec{p}_{sup}(p=0.7,\lambda=0.2)$.}
    \label{subfig:addtion_high_lamb}
    \end{subfigure}
    \begin{subfigure}{\linewidth}
    \includegraphics[width=\textwidth]{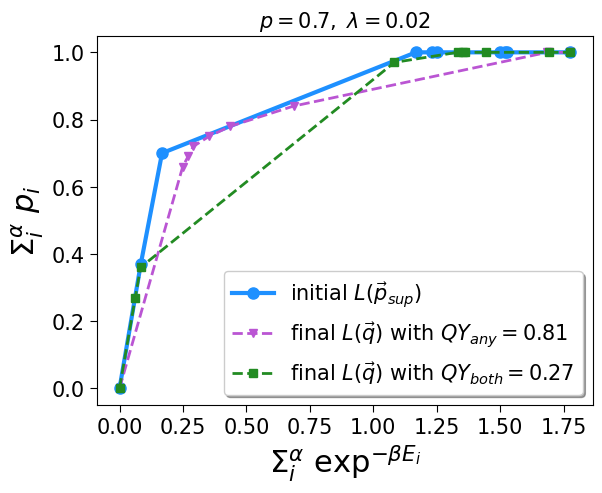}
    \caption{\textbf{Low coherence}, with initial state: $\vec{p}_{sup}(p=0.7,\lambda=0.02)$.}
    \label{subfig:addtion_low_lamb}
    \end{subfigure}
    \caption{\textbf{Effect of coherences on quantum yield.} The figures show an initial state (blue) and below its final states that maximise the quantum yield whether it is defined as $QY_{any}$ (purple) or $QY_{both}$ (green dotted). Higher coherence (fig.~\ref{subfig:addtion_high_lamb}) leads to higher quantum yields, see $QY_{both}$.}
    \label{fig:superposition}
\end{figure}

The increase in coherence leads to higher quantum yields, see fig.~\ref{fig:superposition}. The higher the initial Lorenz curve, the more states are thermomajorized by it so it can allow higher final occupation probabilities. Indeed $QY_{both}=0.39$ when $\lambda=0.2$ (fig.~\ref{subfig:addtion_high_lamb}) but only $QY_{both}=0.27$ when $\lambda=0.02$ (fig.~\ref{subfig:addtion_low_lamb}). There is a significant net increase from 27\% to 39\%. 
The Lorenz curves of these final optimal states are shown in green dotted. They are expectedly both below the initial blue line, because the final states are thermomajorized by $\Vec{p}_{sup}(p,\lambda)$.

Let us consider the role of coherence in this increase from 27\% to 39\%. If one tries to find the final state with the best quantum yield $QY_{both}$, by definition one has to construct the optimal Lorenz curve that is below the initial blue curve in figure~\ref{fig:superposition} and that maximises $q_{tt}$. A maximal $q_{tt}$ will likely end up in first place in the reordering of elements (see eq.~(\ref{eq:reordering})). This gives us the first data point (after the origin) of $L(\Vec{q})$: $(e^{-\beta E_{tt}},q_{tt})$. It has to be below the blue line, because $L(\vec{p}_i)\geq L(\vec{q})$. So the amplitude of $q_{tt}$ is directly impacted by the initial Lorenz curve. This can be seen in figure~\ref{fig:superposition} following the green dotted lines. In figure~\ref{subfig:addtion_low_lamb} the first point is at $0.27$ on the vertical axis, which is the highest possible point below the blue curve for $\text{x}=e^{-\beta E_{tt}}$. This value indeed corresponds to the quantum yield $QY_{both}=0.27$ (and the same for figure~\ref{subfig:addtion_high_lamb} where $QY_{both}=0.39$). 

To summarize, the higher the initial coherence, the higher the initial Lorenz curve, the more possibilities for the final curve and the higher its quantum yield.

The figure also presents the other possible definition of the quantum yield. For $\lambda=0.02$ and for $\lambda=0.2$ the optimal is $QY_{any}=0.81$. The increase in coherence does not seem to affect this definition of the quantum yield at first sight, but this is not true for all energy values.

\begin{figure}[h]
    \centering
    \includegraphics[width=\linewidth]{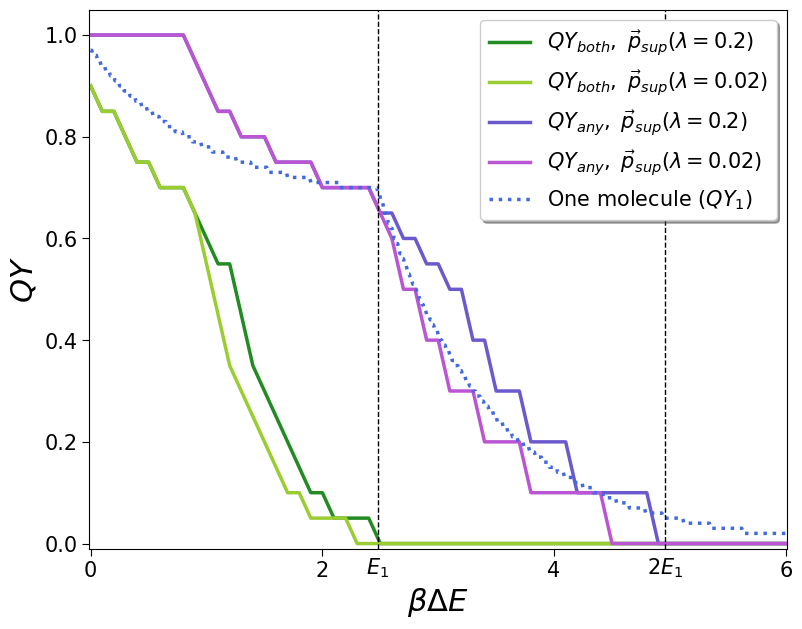}
    \caption{\textbf{Quantum yields as a function of energy gap by superposing two excited molecules.} Different definitions of quantum yield: $QY_{both}$ (green) and $QY_{any}$ (purple). Initial $\Vec{p}_{sup}$ as in eq.(\ref{eq:pop_vec-superposition}). All curves plotted with $p=0.7$. Dark colours represent higher initial coherence ($\lambda=0.2$), lighter colours have lower coherence ($\lambda=0.02$) (accordingly they have lower quantum yields). For comparison, the quantum yield for a single molecule $QY_1$ (blue dotted).}
    \label{fig:delta_superposition}
\end{figure}

In figure~\ref{fig:delta_superposition} we show the variation of the quantum yield as a function of energy difference, highlighting the best possible quantum yield one can obtain with thermomajorization for different energy gaps $\beta \Delta E$. The darker colours show results where the initial state corresponds to $\lambda=0.2$, and the lighter ones to $\lambda=0.02$. The purple curves correspond to the $QY_{any}$ definition of the quantum yield, while the green ones show the optimal $QY_{both}$. In both cases, the darker line remains above the lighter one, because higher coherence induces a higher optimal quantum yield independently of which definition we choose.
A hierarchy exists among these definitions, since $QY_{any}$ contains more terms, it will always be higher than $QY_{both}$.

The curves appear step-like due to the sampling resolution along the vertical axis resulting from high computational overhead. The simulation evaluates all possible population vectors in a vector space with dimension $3^2=9$ and checks whether they have the highest quantum yield, and are thermomajorized by the initial state. 

The green dotted horizontal lines show when the energy gap $\Delta E$ equals $E_1$ and $2E_1$. These energies are related to states where one molecule is excited, or both, respectively. The lines correspond to the cusp where $QY_{any}$ begin to decrease. When the energy gap is larger than the initial excitations it becomes harder to reach the \textit{trans} state. So there is an exponential decrease of the quantum yield with increasing $\Delta E$. At $\Delta E > E_1$, states with a single excited molecule can no longer access to the \textit{trans} state, so now the quantum yield takes
lower values. Finally, when $\Delta E$ exceeds double the excitation energy, it becomes impossible for any initial state to reach the \textit{trans} state and the quantum yield tends to zero.

Figure~\ref{fig:delta_superposition} also provides a comparison to the quantum yield of one molecule (with p=0.7). It is higher than 
$QY_{both}$ (in green) regardless the initial coherence, because photoisomerizing both molecules takes twice the energy. Indeed, the green curves already start decreasing towards 0 at $\Delta E\approx E_1$ instead of $2E_1$ like the purple ones. 
Notably for $QY_{any}$, higher coherence levels become crucial for surpassing the $QY_{1}$ threshold, as in the case of the dark purple line, which is above the blue dotted one. This highlights the importance of enhanced coherence in achieving higher quantum yields.

\begin{figure}[h]
    \centering
    \includegraphics[width=\linewidth]{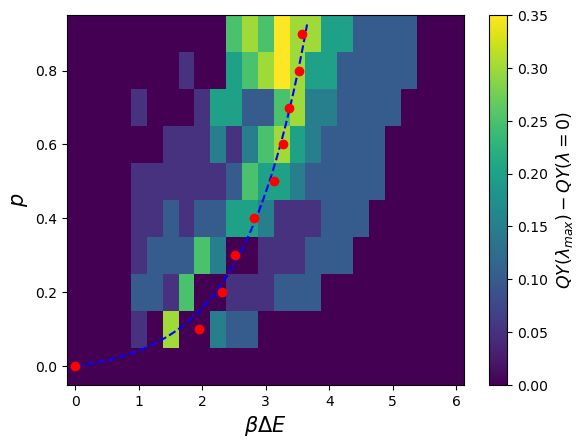}
    \caption{\textbf{The increase of quantum yield $QY_{any}$ as a function of excitation probability $p$ and $\beta \Delta E$.} It corresponds to the difference of $QY_{any}$ between the maximally coherent and noncoherent case. The red points indicate for each $p$ the average $\Delta E$ where the difference is maximal. The blue dotted line is the fitted curve $p=f(\beta\Delta E)=p_0(e^{\beta\Delta E}-1)$ passing through these points.}
    \label{fig:2D Map}
\end{figure}
Interestingly, coherence does not have the same effect for all values of $\Delta E$. To illustrate the intricate dependency on multiple parameters, figure~\ref{fig:2D Map} shows the increase of the quantum yield $QY_{any}$ as a function of both the initial probability $p$ and the energy gap $\Delta E$. The plotted quantity is the difference between the case where the states are maximally coherent, $\lambda_{max}=p/2$, and the noncoherent case, $\lambda=0$: $QY_{any}(\lambda_{max})-QY_{any}(\lambda=0)$. Due to the important computational overhead, the figure is coarse-grained. However one can clearly distinguish a line $p=f(\beta\Delta E)$ where the effect of coherence is the strongest. The average points of maximal effect (in red) seem to fit an exponential curve $f(\beta\Delta E)=p_0(e^{\beta\Delta E}-1)$ (in blue) where $p_0\approx0.025$. It will be the task of a subsequent work to analyse this dependence exactly, and to increase the precision of the numerical results.

The results in this study were obtained for $\beta=1\text{eV}^{-1}$ corresponding to unphysical temperatures limiting physical insight into photoisomerization. If one works at room temperature, the exponential decrease of figure \ref{fig:delta_superposition} becomes very sharp, because the energy levels are too far apart compared to the thermal excitations. In other words, the thermal bath cannot provide sufficient energy to boost
transitions between states. This leads to a quantum yield limit that can only be reached from higher excited states.

\section{CONCLUSION}
\label{sec:conclusion}

In general, quantum thermodynamical models avoid the high temperature limit. However, this work finds that higher temperatures enable transitions to higher-energy states that are inaccessible at lower temperatures.
Real systems are more complicated. The discrete energy levels considered here correspond in reality to broad continuous absorption peaks. We neglect also the environment so an open systems approach might be necessary to understand the specificity of complex systems to quantify the impact of coherence and thermomajorization on its photoisomerization yield. A particularly interesting case-study is Rhodopsin, which benefits from advanced simulation techniques \cite{yang_quantumclassical_2022,pedraza-gonzalez_evolution_2022} and is of wide interest for biochemical applications \cite{kurihara_microbial_2015,kojima_unlimited_2020,kandori_retinal_2020}. There are hints of the importance coherence during the photoisomerizing process~\cite{prokhorenko_coherent_2006,wang_vibrationally_1994}. Our methodology of considering the role of coherence and thermomajorization could in principle help to estimate the efficiency of photoisomerization and to gain a deeper understanding of its drivers. In future works it could be interesting to relate coherence to other types of resources like classical and quantum correlations.

\section{Acknowledgements}
We thank Prof. Massimo Olivucci and Dr Laura Pedraza-González for helpful discussions on their simulations of rhodopsin and for sharing simulated parameters. T.F. thanks the Gordon and Betty Moore Foundation, Lillian Martin and the Oxford Martin School, and the John Fell Fund for support. O.P. acknowledges support by the Scientific and Technological Research Council of Turkey (TUBITAK) under Grant No. 120F089. M.B. thanks the ENS Paris-Saclay ARPE programme for support.

\bibliography{bibliography.bib}

\begin{thebibliography}{42}%
\makeatletter
\providecommand \@ifxundefined [1]{%
 \@ifx{#1\undefined}
}%
\providecommand \@ifnum [1]{%
 \ifnum #1\expandafter \@firstoftwo
 \else \expandafter \@secondoftwo
 \fi
}%
\providecommand \@ifx [1]{%
 \ifx #1\expandafter \@firstoftwo
 \else \expandafter \@secondoftwo
 \fi
}%
\providecommand \natexlab [1]{#1}%
\providecommand \enquote  [1]{``#1''}%
\providecommand \bibnamefont  [1]{#1}%
\providecommand \bibfnamefont [1]{#1}%
\providecommand \citenamefont [1]{#1}%
\providecommand \href@noop [0]{\@secondoftwo}%
\providecommand \href [0]{\begingroup \@sanitize@url \@href}%
\providecommand \@href[1]{\@@startlink{#1}\@@href}%
\providecommand \@@href[1]{\endgroup#1\@@endlink}%
\providecommand \@sanitize@url [0]{\catcode `\\12\catcode `\$12\catcode `\&12\catcode `\#12\catcode `\^12\catcode `\_12\catcode `\%12\relax}%
\providecommand \@@startlink[1]{}%
\providecommand \@@endlink[0]{}%
\providecommand \url  [0]{\begingroup\@sanitize@url \@url }%
\providecommand \@url [1]{\endgroup\@href {#1}{\urlprefix }}%
\providecommand \urlprefix  [0]{URL }%
\providecommand \Eprint [0]{\href }%
\providecommand \doibase [0]{https://doi.org/}%
\providecommand \selectlanguage [0]{\@gobble}%
\providecommand \bibinfo  [0]{\@secondoftwo}%
\providecommand \bibfield  [0]{\@secondoftwo}%
\providecommand \translation [1]{[#1]}%
\providecommand \BibitemOpen [0]{}%
\providecommand \bibitemStop [0]{}%
\providecommand \bibitemNoStop [0]{.\EOS\space}%
\providecommand \EOS [0]{\spacefactor3000\relax}%
\providecommand \BibitemShut  [1]{\csname bibitem#1\endcsname}%
\let\auto@bib@innerbib\@empty
\bibitem [{\citenamefont {Goold}\ \emph {et~al.}(2016)\citenamefont {Goold}, \citenamefont {Huber}, \citenamefont {Riera}, \citenamefont {Rio},\ and\ \citenamefont {Skrzypczyk}}]{goold_role_2016}%
  \BibitemOpen
  \bibfield  {author} {\bibinfo {author} {\bibfnamefont {J.}~\bibnamefont {Goold}}, \bibinfo {author} {\bibfnamefont {M.}~\bibnamefont {Huber}}, \bibinfo {author} {\bibfnamefont {A.}~\bibnamefont {Riera}}, \bibinfo {author} {\bibfnamefont {L.~d.}\ \bibnamefont {Rio}},\ and\ \bibinfo {author} {\bibfnamefont {P.}~\bibnamefont {Skrzypczyk}},\ }\href {https://doi.org/10.1088/1751-8113/49/14/143001} {\bibfield  {journal} {\bibinfo  {journal} {Journal of Physics A: Mathematical and Theoretical}\ }\textbf {\bibinfo {volume} {49}},\ \bibinfo {pages} {143001} (\bibinfo {year} {2016})},\ \bibinfo {note} {publisher: IOP Publishing}\BibitemShut {NoStop}%
\bibitem [{\citenamefont {Vinjanampathy}\ and\ \citenamefont {Anders}(2016)}]{Vinjanampathy2016}%
  \BibitemOpen
  \bibfield  {author} {\bibinfo {author} {\bibfnamefont {S.}~\bibnamefont {Vinjanampathy}}\ and\ \bibinfo {author} {\bibfnamefont {J.}~\bibnamefont {Anders}},\ }\href {https://doi.org/10.1080/00107514.2016.1201896} {\bibfield  {journal} {\bibinfo  {journal} {Contemporary Physics}\ }\textbf {\bibinfo {volume} {57}},\ \bibinfo {pages} {545} (\bibinfo {year} {2016})}\BibitemShut {NoStop}%
\bibitem [{\citenamefont {Binder}\ \emph {et~al.}(2018)\citenamefont {Binder}, \citenamefont {Correa}, \citenamefont {Gogolin}, \citenamefont {Anders},\ and\ \citenamefont {Adesso}}]{QTD_2018}%
  \BibitemOpen
  \bibinfo {editor} {\bibfnamefont {F.}~\bibnamefont {Binder}}, \bibinfo {editor} {\bibfnamefont {L.~A.}\ \bibnamefont {Correa}}, \bibinfo {editor} {\bibfnamefont {C.}~\bibnamefont {Gogolin}}, \bibinfo {editor} {\bibfnamefont {J.}~\bibnamefont {Anders}},\ and\ \bibinfo {editor} {\bibfnamefont {G.}~\bibnamefont {Adesso}},\ eds.,\ \href {https://doi.org/10.1007/978-3-319-99046-0} {\emph {\bibinfo {title} {Thermodynamics in the Quantum Regime: Fundamental Aspects and New Directions}}}\ (\bibinfo  {publisher} {Springer International Publishing},\ \bibinfo {year} {2018})\BibitemShut {NoStop}%
\bibitem [{\citenamefont {Deffner}\ and\ \citenamefont {Campbell}(2019)}]{QTD_2019}%
  \BibitemOpen
  \bibfield  {author} {\bibinfo {author} {\bibfnamefont {S.}~\bibnamefont {Deffner}}\ and\ \bibinfo {author} {\bibfnamefont {S.}~\bibnamefont {Campbell}},\ }\href {https://doi.org/10.1088/2053-2571/ab21c6} {\emph {\bibinfo {title} {Quantum Thermodynamics}}},\ 2053-2571\ (\bibinfo  {publisher} {Morgan {\&} Claypool Publishers},\ \bibinfo {year} {2019})\BibitemShut {NoStop}%
\bibitem [{\citenamefont {Alicki}\ and\ \citenamefont {Kosloff}(2018)}]{Alicki2018}%
  \BibitemOpen
  \bibfield  {author} {\bibinfo {author} {\bibfnamefont {R.}~\bibnamefont {Alicki}}\ and\ \bibinfo {author} {\bibfnamefont {R.}~\bibnamefont {Kosloff}},\ }\bibinfo {title} {Introduction to quantum thermodynamics: History and prospects},\ in\ \href {https://doi.org/10.1007/978-3-319-99046-0_1} {\emph {\bibinfo {booktitle} {Thermodynamics in the Quantum Regime: Fundamental Aspects and New Directions}}},\ \bibinfo {editor} {edited by\ \bibinfo {editor} {\bibfnamefont {F.}~\bibnamefont {Binder}}, \bibinfo {editor} {\bibfnamefont {L.~A.}\ \bibnamefont {Correa}}, \bibinfo {editor} {\bibfnamefont {C.}~\bibnamefont {Gogolin}}, \bibinfo {editor} {\bibfnamefont {J.}~\bibnamefont {Anders}},\ and\ \bibinfo {editor} {\bibfnamefont {G.}~\bibnamefont {Adesso}}}\ (\bibinfo  {publisher} {Springer International Publishing},\ \bibinfo {address} {Cham},\ \bibinfo {year} {2018})\ pp.\ \bibinfo {pages} {1--33}\BibitemShut {NoStop}%
\bibitem [{\citenamefont {Breuer}\ and\ \citenamefont {Petruccione}(2002)}]{BreuerAndPetruccione-2002}%
  \BibitemOpen
  \bibfield  {author} {\bibinfo {author} {\bibfnamefont {H.~P.}\ \bibnamefont {Breuer}}\ and\ \bibinfo {author} {\bibfnamefont {F.}~\bibnamefont {Petruccione}},\ }\href {https://doi.org/10.1093/acprof:oso/9780199213900.001.0001} {\emph {\bibinfo {title} {The theory of open quantum systems}}},\ \bibinfo {edition} {1st}\ ed.\ (\bibinfo  {publisher} {Oxford University Press},\ \bibinfo {year} {2002})\ Chap.~\bibinfo {chapter} {3}, pp.\ \bibinfo {pages} {130--137}\BibitemShut {NoStop}%
\bibitem [{\citenamefont {Wootters}\ and\ \citenamefont {Zurek}(1982)}]{Wootters1982}%
  \BibitemOpen
  \bibfield  {author} {\bibinfo {author} {\bibfnamefont {W.~K.}\ \bibnamefont {Wootters}}\ and\ \bibinfo {author} {\bibfnamefont {W.~H.}\ \bibnamefont {Zurek}},\ }\href {https://doi.org/10.1038/299802a0} {\bibfield  {journal} {\bibinfo  {journal} {Nature}\ }\textbf {\bibinfo {volume} {299}},\ \bibinfo {pages} {802} (\bibinfo {year} {1982})}\BibitemShut {NoStop}%
\bibitem [{\citenamefont {Gour}\ \emph {et~al.}(2015)\citenamefont {Gour}, \citenamefont {M\"{u}ller}, \citenamefont {Narasimhachar}, \citenamefont {Spekkens},\ and\ \citenamefont {Halpern}}]{2015_PhysReport_Nicole_Review}%
  \BibitemOpen
  \bibfield  {author} {\bibinfo {author} {\bibfnamefont {G.}~\bibnamefont {Gour}}, \bibinfo {author} {\bibfnamefont {M.~P.}\ \bibnamefont {M\"{u}ller}}, \bibinfo {author} {\bibfnamefont {V.}~\bibnamefont {Narasimhachar}}, \bibinfo {author} {\bibfnamefont {R.~W.}\ \bibnamefont {Spekkens}},\ and\ \bibinfo {author} {\bibfnamefont {N.~Y.}\ \bibnamefont {Halpern}},\ }\href {https://doi.org/https://doi.org/10.1016/j.physrep.2015.04.003} {\bibfield  {journal} {\bibinfo  {journal} {\emph{Physics Reports}}\ }\textbf {\bibinfo {volume} {583}},\ \bibinfo {pages} {1} (\bibinfo {year} {2015})}\BibitemShut {NoStop}%
\bibitem [{\citenamefont {Ng}\ and\ \citenamefont {Woods}(2018)}]{ng_resource_2018}%
  \BibitemOpen
  \bibfield  {author} {\bibinfo {author} {\bibfnamefont {N.~H.~Y.}\ \bibnamefont {Ng}}\ and\ \bibinfo {author} {\bibfnamefont {M.~P.}\ \bibnamefont {Woods}},\ }\bibinfo {title} {Resource theory of quantum thermodynamics: Thermal operations and second laws},\ in\ \href {https://doi.org/10.1007/978-3-319-99046-0_26} {\emph {\bibinfo {booktitle} {Thermodynamics in the Quantum Regime: Fundamental Aspects and New Directions}}},\ \bibinfo {editor} {edited by\ \bibinfo {editor} {\bibfnamefont {F.}~\bibnamefont {Binder}}, \bibinfo {editor} {\bibfnamefont {L.~A.}\ \bibnamefont {Correa}}, \bibinfo {editor} {\bibfnamefont {C.}~\bibnamefont {Gogolin}}, \bibinfo {editor} {\bibfnamefont {J.}~\bibnamefont {Anders}},\ and\ \bibinfo {editor} {\bibfnamefont {G.}~\bibnamefont {Adesso}}}\ (\bibinfo  {publisher} {Springer International Publishing},\ \bibinfo {address} {Cham},\ \bibinfo {year} {2018})\ pp.\ \bibinfo {pages} {625--650}\BibitemShut {NoStop}%
\bibitem [{\citenamefont {Lostaglio}(2019)}]{lostaglio_introductory_2019}%
  \BibitemOpen
  \bibfield  {author} {\bibinfo {author} {\bibfnamefont {M.}~\bibnamefont {Lostaglio}},\ }\href {https://doi.org/10.1088/1361-6633/ab46e5} {\bibfield  {journal} {\bibinfo  {journal} {Reports on Progress in Physics}\ }\textbf {\bibinfo {volume} {82}},\ \bibinfo {pages} {114001} (\bibinfo {year} {2019})},\ \bibinfo {note} {publisher: IOP Publishing}\BibitemShut {NoStop}%
\bibitem [{\citenamefont {Torun}\ \emph {et~al.}(2023)\citenamefont {Torun}, \citenamefont {Pusuluk},\ and\ \citenamefont {M\"{u}stecapl{\i}o\u{g}lu}}]{TORUN2023}%
  \BibitemOpen
  \bibfield  {author} {\bibinfo {author} {\bibfnamefont {G.}~\bibnamefont {Torun}}, \bibinfo {author} {\bibfnamefont {O.}~\bibnamefont {Pusuluk}},\ and\ \bibinfo {author} {\bibfnamefont {O.~E.}\ \bibnamefont {M\"{u}stecapl{\i}o\u{g}lu}},\ }\href {https://doi.org/10.55730/1300-0101.2744} {\bibfield  {journal} {\bibinfo  {journal} {Turkish Journal of Physics}\ }\textbf {\bibinfo {volume} {47}},\ \bibinfo {pages} {141} (\bibinfo {year} {2023})}\BibitemShut {NoStop}%
\bibitem [{\citenamefont {Coecke}\ \emph {et~al.}(2016)\citenamefont {Coecke}, \citenamefont {Fritz},\ and\ \citenamefont {Spekkens}}]{coecke_mathematical_2016}%
  \BibitemOpen
  \bibfield  {author} {\bibinfo {author} {\bibfnamefont {B.}~\bibnamefont {Coecke}}, \bibinfo {author} {\bibfnamefont {T.}~\bibnamefont {Fritz}},\ and\ \bibinfo {author} {\bibfnamefont {R.~W.}\ \bibnamefont {Spekkens}},\ }\href {https://doi.org/10.1016/j.ic.2016.02.008} {\bibfield  {journal} {\bibinfo  {journal} {Information and Computation}\ }\bibinfo {series} {Quantum {Physics} and {Logic}},\ \textbf {\bibinfo {volume} {250}},\ \bibinfo {pages} {59} (\bibinfo {year} {2016})}\BibitemShut {NoStop}%
\bibitem [{\citenamefont {Chitambar}\ and\ \citenamefont {Gour}(2019)}]{chitambar_quantum_2019}%
  \BibitemOpen
  \bibfield  {author} {\bibinfo {author} {\bibfnamefont {E.}~\bibnamefont {Chitambar}}\ and\ \bibinfo {author} {\bibfnamefont {G.}~\bibnamefont {Gour}},\ }\bibfield  {journal} {\bibinfo  {journal} {Reviews of Modern Physics}\ }\textbf {\bibinfo {volume} {91}},\ \href {https://doi.org/10.1103/RevModPhys.91.025001} {10.1103/RevModPhys.91.025001} (\bibinfo {year} {2019})\BibitemShut {NoStop}%
\bibitem [{\citenamefont {Brask}\ \emph {et~al.}(2015)\citenamefont {Brask}, \citenamefont {Haack}, \citenamefont {Brunner},\ and\ \citenamefont {Huber}}]{2015_NewJPhys_Brunner_T2Ent}%
  \BibitemOpen
  \bibfield  {author} {\bibinfo {author} {\bibfnamefont {J.~B.}\ \bibnamefont {Brask}}, \bibinfo {author} {\bibfnamefont {G.}~\bibnamefont {Haack}}, \bibinfo {author} {\bibfnamefont {N.}~\bibnamefont {Brunner}},\ and\ \bibinfo {author} {\bibfnamefont {M.}~\bibnamefont {Huber}},\ }\href {https://doi.org/10.1088/1367-2630/17/11/113029} {\bibfield  {journal} {\bibinfo  {journal} {\emph{New Journal of Physics}}\ }\textbf {\bibinfo {volume} {17}},\ \bibinfo {pages} {113029} (\bibinfo {year} {2015})}\BibitemShut {NoStop}%
\bibitem [{\citenamefont {\ifmmode~\mbox{\c{C}}\else \c{C}\fi{}akmak}\ \emph {et~al.}(2017)\citenamefont {\ifmmode~\mbox{\c{C}}\else \c{C}\fi{}akmak}, \citenamefont {Manatuly},\ and\ \citenamefont {M\"{u}stecapl{\i}o\u{g}lu}}]{2017_PRA_Ozgur}%
  \BibitemOpen
  \bibfield  {author} {\bibinfo {author} {\bibfnamefont {B.}~\bibnamefont {\ifmmode~\mbox{\c{C}}\else \c{C}\fi{}akmak}}, \bibinfo {author} {\bibfnamefont {A.}~\bibnamefont {Manatuly}},\ and\ \bibinfo {author} {\bibfnamefont {O.~E.}\ \bibnamefont {M\"{u}stecapl{\i}o\u{g}lu}},\ }\href {https://doi.org/10.1103/PhysRevA.96.032117} {\bibfield  {journal} {\bibinfo  {journal} {\emph{Physical Review A}}\ }\textbf {\bibinfo {volume} {96}},\ \bibinfo {pages} {032117} (\bibinfo {year} {2017})}\BibitemShut {NoStop}%
\bibitem [{\citenamefont {Henao}\ and\ \citenamefont {Serra}(2018)}]{2018_PRE_RoleOfQCohInHT}%
  \BibitemOpen
  \bibfield  {author} {\bibinfo {author} {\bibfnamefont {I.}~\bibnamefont {Henao}}\ and\ \bibinfo {author} {\bibfnamefont {R.~M.}\ \bibnamefont {Serra}},\ }\href {https://doi.org/10.1103/PhysRevE.97.062105} {\bibfield  {journal} {\bibinfo  {journal} {\emph{Physical Review E}}\ }\textbf {\bibinfo {volume} {97}},\ \bibinfo {pages} {062105} (\bibinfo {year} {2018})}\BibitemShut {NoStop}%
\bibitem [{\citenamefont {Manzano}\ \emph {et~al.}(2019)\citenamefont {Manzano}, \citenamefont {Silva},\ and\ \citenamefont {Parrondo}}]{2019_PRE_T2QCoh}%
  \BibitemOpen
  \bibfield  {author} {\bibinfo {author} {\bibfnamefont {G.}~\bibnamefont {Manzano}}, \bibinfo {author} {\bibfnamefont {R.}~\bibnamefont {Silva}},\ and\ \bibinfo {author} {\bibfnamefont {J.~M.~R.}\ \bibnamefont {Parrondo}},\ }\href {https://doi.org/10.1103/PhysRevE.99.042135} {\bibfield  {journal} {\bibinfo  {journal} {\emph{Physical Review E}}\ }\textbf {\bibinfo {volume} {99}},\ \bibinfo {pages} {042135} (\bibinfo {year} {2019})}\BibitemShut {NoStop}%
\bibitem [{\citenamefont {Tavakoli}\ \emph {et~al.}(2020)\citenamefont {Tavakoli}, \citenamefont {Haack}, \citenamefont {Brunner},\ and\ \citenamefont {Brask}}]{2020_PRA_Brunner_AutonomousMultipartiteEntEngines}%
  \BibitemOpen
  \bibfield  {author} {\bibinfo {author} {\bibfnamefont {A.}~\bibnamefont {Tavakoli}}, \bibinfo {author} {\bibfnamefont {G.}~\bibnamefont {Haack}}, \bibinfo {author} {\bibfnamefont {N.}~\bibnamefont {Brunner}},\ and\ \bibinfo {author} {\bibfnamefont {J.~B.}\ \bibnamefont {Brask}},\ }\href {https://doi.org/10.1103/PhysRevA.101.012315} {\bibfield  {journal} {\bibinfo  {journal} {\emph{Physical Review A}}\ }\textbf {\bibinfo {volume} {101}},\ \bibinfo {pages} {012315} (\bibinfo {year} {2020})}\BibitemShut {NoStop}%
\bibitem [{\citenamefont {Tuncer}\ and\ \citenamefont {M\"{u}stecapl{\i}o\u{g}lu}(2020)}]{2020_AsliReview}%
  \BibitemOpen
  \bibfield  {author} {\bibinfo {author} {\bibfnamefont {A.}~\bibnamefont {Tuncer}}\ and\ \bibinfo {author} {\bibfnamefont {O.}~\bibnamefont {M\"{u}stecapl{\i}o\u{g}lu}},\ }\href {https://doi.org/10.3906/fiz-2009-12} {\bibfield  {journal} {\bibinfo  {journal} {Turkish Journal Physics}\ }\textbf {\bibinfo {volume} {44}},\ \bibinfo {pages} {404} (\bibinfo {year} {2020})}\BibitemShut {NoStop}%
\bibitem [{\citenamefont {Da\u{g}}\ \emph {et~al.}(2016)\citenamefont {Da\u{g}}, \citenamefont {Niedenzu}, \citenamefont {M\"{u}stecapl{\i}o\u{g}lu},\ and\ \citenamefont {Kurizki}}]{2016_Entropy_Ozgur}%
  \BibitemOpen
  \bibfield  {author} {\bibinfo {author} {\bibfnamefont {C.}~\bibnamefont {Da\u{g}}}, \bibinfo {author} {\bibfnamefont {W.}~\bibnamefont {Niedenzu}}, \bibinfo {author} {\bibfnamefont {O.~E.}\ \bibnamefont {M\"{u}stecapl{\i}o\u{g}lu}},\ and\ \bibinfo {author} {\bibfnamefont {G.}~\bibnamefont {Kurizki}},\ }\href {https://doi.org/10.3390/e18070244} {\bibfield  {journal} {\bibinfo  {journal} {\emph{Entropy}}\ }\textbf {\bibinfo {volume} {18}},\ \bibinfo {pages} {244} (\bibinfo {year} {2016})}\BibitemShut {NoStop}%
\bibitem [{\citenamefont {Manatuly}\ \emph {et~al.}(2019)\citenamefont {Manatuly}, \citenamefont {Niedenzu}, \citenamefont {Rom\'an-Ancheyta}, \citenamefont {\ifmmode~\mbox{\c{C}}\else \c{C}\fi{}akmak}, \citenamefont {M\"ustecapl\ifmmode \imath \else \i \fi{}o\ifmmode~\breve{g}\else \u{g}\fi{}lu},\ and\ \citenamefont {Kurizki}}]{2019_PRE_Ozgur_Multiatom}%
  \BibitemOpen
  \bibfield  {author} {\bibinfo {author} {\bibfnamefont {A.}~\bibnamefont {Manatuly}}, \bibinfo {author} {\bibfnamefont {W.}~\bibnamefont {Niedenzu}}, \bibinfo {author} {\bibfnamefont {R.}~\bibnamefont {Rom\'an-Ancheyta}}, \bibinfo {author} {\bibfnamefont {B.}~\bibnamefont {\ifmmode~\mbox{\c{C}}\else \c{C}\fi{}akmak}}, \bibinfo {author} {\bibfnamefont {O.~E.}\ \bibnamefont {M\"ustecapl\ifmmode \imath \else \i \fi{}o\ifmmode~\breve{g}\else \u{g}\fi{}lu}},\ and\ \bibinfo {author} {\bibfnamefont {G.}~\bibnamefont {Kurizki}},\ }\href {https://doi.org/10.1103/PhysRevE.99.042145} {\bibfield  {journal} {\bibinfo  {journal} {Physical Review E}\ }\textbf {\bibinfo {volume} {99}},\ \bibinfo {pages} {042145} (\bibinfo {year} {2019})}\BibitemShut {NoStop}%
\bibitem [{\citenamefont {Latune}\ \emph {et~al.}(2019{\natexlab{a}})\citenamefont {Latune}, \citenamefont {Sinayskiy},\ and\ \citenamefont {Petruccione}}]{AHF_2019_PhysRevResearch_Petruccione}%
  \BibitemOpen
  \bibfield  {author} {\bibinfo {author} {\bibfnamefont {C.~L.}\ \bibnamefont {Latune}}, \bibinfo {author} {\bibfnamefont {I.}~\bibnamefont {Sinayskiy}},\ and\ \bibinfo {author} {\bibfnamefont {F.}~\bibnamefont {Petruccione}},\ }\href {https://doi.org/10.1103/PhysRevResearch.1.033097} {\bibfield  {journal} {\bibinfo  {journal} {Physical Review Research}\ }\textbf {\bibinfo {volume} {1}},\ \bibinfo {pages} {033097} (\bibinfo {year} {2019}{\natexlab{a}})}\BibitemShut {NoStop}%
\bibitem [{\citenamefont {Latune}\ \emph {et~al.}(2019{\natexlab{b}})\citenamefont {Latune}, \citenamefont {Sinayskiy},\ and\ \citenamefont {Petruccione}}]{AutoThermalMach_2019}%
  \BibitemOpen
  \bibfield  {author} {\bibinfo {author} {\bibfnamefont {C.~L.}\ \bibnamefont {Latune}}, \bibinfo {author} {\bibfnamefont {I.}~\bibnamefont {Sinayskiy}},\ and\ \bibinfo {author} {\bibfnamefont {F.}~\bibnamefont {Petruccione}},\ }\href {https://doi.org/10.1038/s41598-019-39300-4} {\bibfield  {journal} {\bibinfo  {journal} {Scientific Reports}\ }\textbf {\bibinfo {volume} {9}},\ \bibinfo {pages} {3191} (\bibinfo {year} {2019}{\natexlab{b}})}\BibitemShut {NoStop}%
\bibitem [{\citenamefont {Latune}\ \emph {et~al.}(2020)\citenamefont {Latune}, \citenamefont {Sinayskiy},\ and\ \citenamefont {Petruccione}}]{2020_PRA_HorizantolCohAndPops}%
  \BibitemOpen
  \bibfield  {author} {\bibinfo {author} {\bibfnamefont {C.~L.}\ \bibnamefont {Latune}}, \bibinfo {author} {\bibfnamefont {I.}~\bibnamefont {Sinayskiy}},\ and\ \bibinfo {author} {\bibfnamefont {F.}~\bibnamefont {Petruccione}},\ }\href {https://doi.org/10.1103/PhysRevA.102.042220} {\bibfield  {journal} {\bibinfo  {journal} {Physical Review A}\ }\textbf {\bibinfo {volume} {102}},\ \bibinfo {pages} {042220} (\bibinfo {year} {2020})}\BibitemShut {NoStop}%
\bibitem [{\citenamefont {Pusuluk}\ and\ \citenamefont {M\"ustecapl\ifmmode \imath \else \i \fi{}o\ifmmode~\breve{g}\else \u{g}\fi{}lu}(2021)}]{pusuluk_quantum_2021}%
  \BibitemOpen
  \bibfield  {author} {\bibinfo {author} {\bibfnamefont {O.}~\bibnamefont {Pusuluk}}\ and\ \bibinfo {author} {\bibfnamefont {O.~E.}\ \bibnamefont {M\"ustecapl\ifmmode \imath \else \i \fi{}o\ifmmode~\breve{g}\else \u{g}\fi{}lu}},\ }\href {https://doi.org/10.1103/PhysRevResearch.3.023235} {\bibfield  {journal} {\bibinfo  {journal} {Physical Review Research}\ }\textbf {\bibinfo {volume} {3}},\ \bibinfo {pages} {023235} (\bibinfo {year} {2021})},\ \bibinfo {note} {publisher: American Physical Society}\BibitemShut {NoStop}%
\bibitem [{\citenamefont {Lostaglio}\ \emph {et~al.}(2015)\citenamefont {Lostaglio}, \citenamefont {Korzekwa}, \citenamefont {Jennings},\ and\ \citenamefont {Rudolph}}]{lostaglio_quantum_2015}%
  \BibitemOpen
  \bibfield  {author} {\bibinfo {author} {\bibfnamefont {M.}~\bibnamefont {Lostaglio}}, \bibinfo {author} {\bibfnamefont {K.}~\bibnamefont {Korzekwa}}, \bibinfo {author} {\bibfnamefont {D.}~\bibnamefont {Jennings}},\ and\ \bibinfo {author} {\bibfnamefont {T.}~\bibnamefont {Rudolph}},\ }\href {https://doi.org/10.1103/PhysRevX.5.021001} {\bibfield  {journal} {\bibinfo  {journal} {Physical Review X}\ }\textbf {\bibinfo {volume} {5}},\ \bibinfo {pages} {021001} (\bibinfo {year} {2015})},\ \bibinfo {note} {publisher: American Physical Society}\BibitemShut {NoStop}%
\bibitem [{\citenamefont {Biswas}\ \emph {et~al.}(2022)\citenamefont {Biswas}, \citenamefont {Junior}, \citenamefont {Horodecki},\ and\ \citenamefont {Korzekwa}}]{biswas_fluctuation-dissipation_2022}%
  \BibitemOpen
  \bibfield  {author} {\bibinfo {author} {\bibfnamefont {T.}~\bibnamefont {Biswas}}, \bibinfo {author} {\bibfnamefont {A.~d.~O.}\ \bibnamefont {Junior}}, \bibinfo {author} {\bibfnamefont {M.}~\bibnamefont {Horodecki}},\ and\ \bibinfo {author} {\bibfnamefont {K.}~\bibnamefont {Korzekwa}},\ }\href {https://doi.org/10.1103/PhysRevE.105.054127} {\bibfield  {journal} {\bibinfo  {journal} {Physical Review E}\ }\textbf {\bibinfo {volume} {105}},\ \bibinfo {pages} {054127} (\bibinfo {year} {2022})},\ \bibinfo {note} {publisher: American Physical Society}\BibitemShut {NoStop}%
\bibitem [{\citenamefont {Yunger~Halpern}\ and\ \citenamefont {Limmer}(2020)}]{yunger_halpern_fundamental_2020}%
  \BibitemOpen
  \bibfield  {author} {\bibinfo {author} {\bibfnamefont {N.}~\bibnamefont {Yunger~Halpern}}\ and\ \bibinfo {author} {\bibfnamefont {D.~T.}\ \bibnamefont {Limmer}},\ }\href {https://doi.org/10.1103/PhysRevA.101.042116} {\bibfield  {journal} {\bibinfo  {journal} {Physical Review A}\ }\textbf {\bibinfo {volume} {101}},\ \bibinfo {pages} {042116} (\bibinfo {year} {2020})},\ \bibinfo {note} {publisher: American Physical Society}\BibitemShut {NoStop}%
\bibitem [{\citenamefont {Spaventa}\ \emph {et~al.}(2022)\citenamefont {Spaventa}, \citenamefont {Huelga},\ and\ \citenamefont {Plenio}}]{spaventa_capacity_2022}%
  \BibitemOpen
  \bibfield  {author} {\bibinfo {author} {\bibfnamefont {G.}~\bibnamefont {Spaventa}}, \bibinfo {author} {\bibfnamefont {S.~F.}\ \bibnamefont {Huelga}},\ and\ \bibinfo {author} {\bibfnamefont {M.~B.}\ \bibnamefont {Plenio}},\ }\href {https://doi.org/10.1103/PhysRevA.105.012420} {\bibfield  {journal} {\bibinfo  {journal} {Physical Review A}\ }\textbf {\bibinfo {volume} {105}},\ \bibinfo {pages} {012420} (\bibinfo {year} {2022})},\ \bibinfo {note} {publisher: American Physical Society}\BibitemShut {NoStop}%
\bibitem [{\citenamefont {Horodecki}\ and\ \citenamefont {Oppenheim}(2013)}]{horodecki_fundamental_2013}%
  \BibitemOpen
  \bibfield  {author} {\bibinfo {author} {\bibfnamefont {M.}~\bibnamefont {Horodecki}}\ and\ \bibinfo {author} {\bibfnamefont {J.}~\bibnamefont {Oppenheim}},\ }\href {https://doi.org/10.1038/ncomms3059} {\bibfield  {journal} {\bibinfo  {journal} {Nature Communications}\ }\textbf {\bibinfo {volume} {4}},\ \bibinfo {pages} {2059} (\bibinfo {year} {2013})},\ \bibinfo {note} {number: 1 Publisher: Nature Publishing Group}\BibitemShut {NoStop}%
\bibitem [{\citenamefont {Ernst}\ \emph {et~al.}(2014)\citenamefont {Ernst}, \citenamefont {Lodowski}, \citenamefont {Elstner}, \citenamefont {Hegemann}, \citenamefont {Brown},\ and\ \citenamefont {Kandori}}]{ernst_microbial_2014}%
  \BibitemOpen
  \bibfield  {author} {\bibinfo {author} {\bibfnamefont {O.~P.}\ \bibnamefont {Ernst}}, \bibinfo {author} {\bibfnamefont {D.~T.}\ \bibnamefont {Lodowski}}, \bibinfo {author} {\bibfnamefont {M.}~\bibnamefont {Elstner}}, \bibinfo {author} {\bibfnamefont {P.}~\bibnamefont {Hegemann}}, \bibinfo {author} {\bibfnamefont {L.~S.}\ \bibnamefont {Brown}},\ and\ \bibinfo {author} {\bibfnamefont {H.}~\bibnamefont {Kandori}},\ }\href {https://doi.org/10.1021/cr4003769} {\bibfield  {journal} {\bibinfo  {journal} {Chemical Reviews}\ }\textbf {\bibinfo {volume} {114}},\ \bibinfo {pages} {126} (\bibinfo {year} {2014})},\ \bibinfo {note} {publisher: American Chemical Society}\BibitemShut {NoStop}%
\bibitem [{\citenamefont {Hahn}\ and\ \citenamefont {Stock}(2000)}]{hahn_quantum-mechanical_2000}%
  \BibitemOpen
  \bibfield  {author} {\bibinfo {author} {\bibfnamefont {S.}~\bibnamefont {Hahn}}\ and\ \bibinfo {author} {\bibfnamefont {G.}~\bibnamefont {Stock}},\ }\href {https://doi.org/10.1021/jp992939g} {\bibfield  {journal} {\bibinfo  {journal} {The Journal of Physical Chemistry B}\ }\textbf {\bibinfo {volume} {104}},\ \bibinfo {pages} {1146} (\bibinfo {year} {2000})},\ \bibinfo {note} {publisher: American Chemical Society}\BibitemShut {NoStop}%
\bibitem [{\citenamefont {Wang}\ \emph {et~al.}(1994)\citenamefont {Wang}, \citenamefont {Schoenlein}, \citenamefont {Peteanu}, \citenamefont {Mathies},\ and\ \citenamefont {Shank}}]{wang_vibrationally_1994}%
  \BibitemOpen
  \bibfield  {author} {\bibinfo {author} {\bibfnamefont {Q.}~\bibnamefont {Wang}}, \bibinfo {author} {\bibfnamefont {R.~W.}\ \bibnamefont {Schoenlein}}, \bibinfo {author} {\bibfnamefont {L.~A.}\ \bibnamefont {Peteanu}}, \bibinfo {author} {\bibfnamefont {R.~A.}\ \bibnamefont {Mathies}},\ and\ \bibinfo {author} {\bibfnamefont {C.~V.}\ \bibnamefont {Shank}},\ }\href {https://doi.org/10.1126/science.7939680} {\bibfield  {journal} {\bibinfo  {journal} {Science}\ }\textbf {\bibinfo {volume} {266}},\ \bibinfo {pages} {422} (\bibinfo {year} {1994})},\ \Eprint {https://arxiv.org/abs/https://www.science.org/doi/pdf/10.1126/science.7939680} {https://www.science.org/doi/pdf/10.1126/science.7939680} \BibitemShut {NoStop}%
\bibitem [{\citenamefont {Pusuluk}\ \emph {et~al.}(2018)\citenamefont {Pusuluk}, \citenamefont {Farrow}, \citenamefont {Deliduman}, \citenamefont {Burnett},\ and\ \citenamefont {Vedral}}]{pusuluk_proton_2018}%
  \BibitemOpen
  \bibfield  {author} {\bibinfo {author} {\bibfnamefont {O.}~\bibnamefont {Pusuluk}}, \bibinfo {author} {\bibfnamefont {T.}~\bibnamefont {Farrow}}, \bibinfo {author} {\bibfnamefont {C.}~\bibnamefont {Deliduman}}, \bibinfo {author} {\bibfnamefont {K.}~\bibnamefont {Burnett}},\ and\ \bibinfo {author} {\bibfnamefont {V.}~\bibnamefont {Vedral}},\ }\href {https://doi.org/10.1098/rspa.2018.0037} {\bibfield  {journal} {\bibinfo  {journal} {Proceedings of the Royal Society A: Mathematical, Physical and Engineering Sciences}\ }\textbf {\bibinfo {volume} {474}},\ \bibinfo {pages} {20180037} (\bibinfo {year} {2018})},\ \bibinfo {note} {publisher: Royal Society}\BibitemShut {NoStop}%
\bibitem [{\citenamefont {Johnson}\ \emph {et~al.}(2017)\citenamefont {Johnson}, \citenamefont {Farag}, \citenamefont {Halpin}, \citenamefont {Morizumi}, \citenamefont {Prokhorenko}, \citenamefont {Knoester}, \citenamefont {Jansen}, \citenamefont {Ernst},\ and\ \citenamefont {Miller}}]{johnson_primary_2017}%
  \BibitemOpen
  \bibfield  {author} {\bibinfo {author} {\bibfnamefont {P.~J.~M.}\ \bibnamefont {Johnson}}, \bibinfo {author} {\bibfnamefont {M.~H.}\ \bibnamefont {Farag}}, \bibinfo {author} {\bibfnamefont {A.}~\bibnamefont {Halpin}}, \bibinfo {author} {\bibfnamefont {T.}~\bibnamefont {Morizumi}}, \bibinfo {author} {\bibfnamefont {V.~I.}\ \bibnamefont {Prokhorenko}}, \bibinfo {author} {\bibfnamefont {J.}~\bibnamefont {Knoester}}, \bibinfo {author} {\bibfnamefont {T.~L.~C.}\ \bibnamefont {Jansen}}, \bibinfo {author} {\bibfnamefont {O.~P.}\ \bibnamefont {Ernst}},\ and\ \bibinfo {author} {\bibfnamefont {R.~J.~D.}\ \bibnamefont {Miller}},\ }\href {https://doi.org/10.1021/acs.jpcb.7b02329} {\bibfield  {journal} {\bibinfo  {journal} {The Journal of Physical Chemistry B}\ }\textbf {\bibinfo {volume} {121}},\ \bibinfo {pages} {4040} (\bibinfo {year} {2017})},\ \bibinfo {note} {pMID: 28358485},\ \Eprint {https://arxiv.org/abs/https://doi.org/10.1021/acs.jpcb.7b02329} {https://doi.org/10.1021/acs.jpcb.7b02329} \BibitemShut {NoStop}%
\bibitem [{\citenamefont {Pedraza-Gonz{\'a}lez}\ \emph {et~al.}(2019)\citenamefont {Pedraza-Gonz{\'a}lez}, \citenamefont {De~Vico}, \citenamefont {Marı{\'i}n}, \citenamefont {Fanelli},\ and\ \citenamefont {Olivucci}}]{pedraza-gonzalez_a-arm_2019}%
  \BibitemOpen
  \bibfield  {author} {\bibinfo {author} {\bibfnamefont {L.}~\bibnamefont {Pedraza-Gonz{\'a}lez}}, \bibinfo {author} {\bibfnamefont {L.}~\bibnamefont {De~Vico}}, \bibinfo {author} {\bibfnamefont {M.~d.~C.}\ \bibnamefont {Marı{\'i}n}}, \bibinfo {author} {\bibfnamefont {F.}~\bibnamefont {Fanelli}},\ and\ \bibinfo {author} {\bibfnamefont {M.}~\bibnamefont {Olivucci}},\ }\href {https://doi.org/10.1021/acs.jctc.9b00061} {\bibfield  {journal} {\bibinfo  {journal} {Journal of Chemical Theory and Computation}\ }\textbf {\bibinfo {volume} {15}},\ \bibinfo {pages} {3134} (\bibinfo {year} {2019})},\ \bibinfo {note} {pMID: 30916955},\ \Eprint {https://arxiv.org/abs/https://doi.org/10.1021/acs.jctc.9b00061} {https://doi.org/10.1021/acs.jctc.9b00061} \BibitemShut {NoStop}%
\bibitem [{\citenamefont {Yang}\ \emph {et~al.}(2022)\citenamefont {Yang}, \citenamefont {Manathunga}, \citenamefont {Gozem}, \citenamefont {Léonard}, \citenamefont {Andruniów},\ and\ \citenamefont {Olivucci}}]{yang_quantumclassical_2022}%
  \BibitemOpen
  \bibfield  {author} {\bibinfo {author} {\bibfnamefont {X.}~\bibnamefont {Yang}}, \bibinfo {author} {\bibfnamefont {M.}~\bibnamefont {Manathunga}}, \bibinfo {author} {\bibfnamefont {S.}~\bibnamefont {Gozem}}, \bibinfo {author} {\bibfnamefont {J.}~\bibnamefont {Léonard}}, \bibinfo {author} {\bibfnamefont {T.}~\bibnamefont {Andruniów}},\ and\ \bibinfo {author} {\bibfnamefont {M.}~\bibnamefont {Olivucci}},\ }\href {https://doi.org/10.1038/s41557-022-00892-6} {\bibfield  {journal} {\bibinfo  {journal} {Nature Chemistry}\ }\textbf {\bibinfo {volume} {14}},\ \bibinfo {pages} {441} (\bibinfo {year} {2022})},\ \bibinfo {note} {number: 4 Publisher: Nature Publishing Group}\BibitemShut {NoStop}%
\bibitem [{\citenamefont {Pedraza-González}\ \emph {et~al.}(2022)\citenamefont {Pedraza-González}, \citenamefont {Barneschi}, \citenamefont {Padula}, \citenamefont {De~Vico},\ and\ \citenamefont {Olivucci}}]{pedraza-gonzalez_evolution_2022}%
  \BibitemOpen
  \bibfield  {author} {\bibinfo {author} {\bibfnamefont {L.}~\bibnamefont {Pedraza-González}}, \bibinfo {author} {\bibfnamefont {L.}~\bibnamefont {Barneschi}}, \bibinfo {author} {\bibfnamefont {D.}~\bibnamefont {Padula}}, \bibinfo {author} {\bibfnamefont {L.}~\bibnamefont {De~Vico}},\ and\ \bibinfo {author} {\bibfnamefont {M.}~\bibnamefont {Olivucci}},\ }\href {https://doi.org/10.1007/s41061-022-00374-w} {\bibfield  {journal} {\bibinfo  {journal} {Topics in Current Chemistry}\ }\textbf {\bibinfo {volume} {380}},\ \bibinfo {pages} {21} (\bibinfo {year} {2022})}\BibitemShut {NoStop}%
\bibitem [{\citenamefont {Kurihara}\ and\ \citenamefont {Sudo}(2015)}]{kurihara_microbial_2015}%
  \BibitemOpen
  \bibfield  {author} {\bibinfo {author} {\bibfnamefont {M.}~\bibnamefont {Kurihara}}\ and\ \bibinfo {author} {\bibfnamefont {Y.}~\bibnamefont {Sudo}},\ }\href {https://doi.org/10.2142/biophysico.12.0_121} {\bibfield  {journal} {\bibinfo  {journal} {Biophysics and Physicobiology}\ }\textbf {\bibinfo {volume} {12}},\ \bibinfo {pages} {121} (\bibinfo {year} {2015})}\BibitemShut {NoStop}%
\bibitem [{\citenamefont {Kojima}\ \emph {et~al.}(2020)\citenamefont {Kojima}, \citenamefont {Shibukawa},\ and\ \citenamefont {Sudo}}]{kojima_unlimited_2020}%
  \BibitemOpen
  \bibfield  {author} {\bibinfo {author} {\bibfnamefont {K.}~\bibnamefont {Kojima}}, \bibinfo {author} {\bibfnamefont {A.}~\bibnamefont {Shibukawa}},\ and\ \bibinfo {author} {\bibfnamefont {Y.}~\bibnamefont {Sudo}},\ }\href {https://doi.org/10.1021/acs.biochem.9b00768} {\bibfield  {journal} {\bibinfo  {journal} {Biochemistry}\ }\textbf {\bibinfo {volume} {59}},\ \bibinfo {pages} {218} (\bibinfo {year} {2020})},\ \bibinfo {note} {publisher: American Chemical Society}\BibitemShut {NoStop}%
\bibitem [{\citenamefont {Kandori}(2020)}]{kandori_retinal_2020}%
  \BibitemOpen
  \bibfield  {author} {\bibinfo {author} {\bibfnamefont {H.}~\bibnamefont {Kandori}},\ }\href {https://doi.org/10.1246/bcsj.20190292} {\bibfield  {journal} {\bibinfo  {journal} {Bulletin of the Chemical Society of Japan}\ }\textbf {\bibinfo {volume} {93}},\ \bibinfo {pages} {76} (\bibinfo {year} {2020})},\ \bibinfo {note} {publisher: The Chemical Society of Japan}\BibitemShut {NoStop}%
\bibitem [{\citenamefont {Prokhorenko}\ \emph {et~al.}(2006)\citenamefont {Prokhorenko}, \citenamefont {Nagy}, \citenamefont {Waschuk}, \citenamefont {Brown}, \citenamefont {Birge},\ and\ \citenamefont {Miller}}]{prokhorenko_coherent_2006}%
  \BibitemOpen
  \bibfield  {author} {\bibinfo {author} {\bibfnamefont {V.~I.}\ \bibnamefont {Prokhorenko}}, \bibinfo {author} {\bibfnamefont {A.~M.}\ \bibnamefont {Nagy}}, \bibinfo {author} {\bibfnamefont {S.~A.}\ \bibnamefont {Waschuk}}, \bibinfo {author} {\bibfnamefont {L.~S.}\ \bibnamefont {Brown}}, \bibinfo {author} {\bibfnamefont {R.~R.}\ \bibnamefont {Birge}},\ and\ \bibinfo {author} {\bibfnamefont {R.~J.~D.}\ \bibnamefont {Miller}},\ }\bibfield  {journal} {\bibinfo  {journal} {Science}\ }\href {https://doi.org/10.1126/science.1130747} {10.1126/science.1130747} (\bibinfo {year} {2006}),\ \bibinfo {note} {publisher: American Association for the Advancement of Science}\BibitemShut {NoStop}%
\end{thebibliography}%

\end{document}